\documentclass[aps,prd,preprint,tightenlines,superscriptaddress,
   showpacs,nofootinbib]{revtex4}
\newcommand{\PRE}[1]{{#1}} % Use if preprint style
\usepackage{bm} \usepackage{epsfig}
\usepackage{multirow}
\newcommand{\be}{\begin{equation}}
\newcommand{\ee}{\end{equation}}
\newcommand{\beq}{\begin{equation}}
\newcommand{\eeq}{\end{equation}}
\newcommand{\beqa}{\begin{eqnarray}}
\newcommand{\eeqa}{\end{eqnarray}}
\newcommand{\no}{\nonumber}

\newcommand{\gev}{\text{GeV}}
\newcommand{\tev}{\text{TeV}}

\newcommand{\eg}{{\em e.g.}}
\newcommand{\ie}{{\em i.e.}}

\newcommand{\eqref}[1]{Eq.~(\ref{#1})}
\newcommand{\eqsref}[2]{Eqs.~(\ref{#1}) and (\ref{#2})}
\newcommand{\secref}[1]{Sec.~\ref{sec:#1}}

\newcommand{\bea}{\begin{eqnarray}}
\newcommand{\eea}{\end{eqnarray}}

\def\slashchar#1{\setbox0=\hbox{$#1$}           % set a box for #1
   \dimen0=\wd0                                 % and get its size
   \setbox1=\hbox{/} \dimen1=\wd1               % get size of /
   \ifdim\dimen0>\dimen1                        % #1 is bigger
      \rlap{\hbox to \dimen0{\hfil/\hfil}}#1
   \else                                        % / is bigger
      \rlap{\hbox to \dimen1{\hfil$#1$\hfil}}/                            
       \fi}
\def\notslashchar#1{\setbox0=\hbox{$#1$}           % set a box for #1
   \dimen0=\wd0                                 % and get its size
   \setbox1=\hbox{/} \dimen1=\wd1               % get size of /
   \ifdim\dimen0>\dimen1                        % #1 is bigger
      \rlap{\hbox to \dimen0{\hfil\phantom{/}\hfil}}#1
   \else                                        % / is bigger
      \rlap{\hbox to \dimen1{\hfil$#1$\hfil}}/                            
       \fi}

\def\ie{{\it i.e.}}
\def\eg{{\it e.g.}}

\begin{document}

\preprint{UCI-TR-2007-49}

\title{ \PRE{\vspace*{1.5in}}The Standard Model and Supersymmetric
  Flavor Puzzles\\
at the Large Hadron Collider \PRE{\vspace*{0.3in}} }

\author{Jonathan L.~Feng}
\affiliation{Department of Physics and Astronomy, University of
California, Irvine, CA 92697, USA }

\author{Christopher G.~Lester}
\affiliation{Cavendish Laboratory, J.~J.~Thomson Avenue, Cambridge,
  CB3 0HE, UK }

\author{Yosef Nir}
\affiliation{Department of Particle Physics, Weizmann Institute of
  Science, Rehovot 76100, Israel }

\author{Yael Shadmi%
\PRE{\vspace*{.2in}} }
\affiliation{Physics Department, Technion--Israel Institute of
Technology, Haifa 32000, Israel \PRE{\vspace*{.5in}} }

\date{December 2007}

\begin{abstract}
\PRE{\vspace*{.3in}} Can the Large Hadron Collider explain the masses
and mixings of the known fermions?  A promising possibility is that
these masses and mixings are determined by flavor symmetries that also
govern new particles that will appear at the LHC.  We consider
well-motivated examples in supersymmetry with both gravity- and
gauge-mediation.  Contrary to spreading belief, new physics need not be
minimally flavor violating.  We build non-minimally flavor violating
models that successfully explain all known lepton masses and mixings,
but span a wide range in their predictions for slepton flavor
violation.  In natural and favorable cases, these models have
metastable sleptons and are characterized by fully reconstructible
events.  We outline many flavor measurements that are then possible
and describe their prospects for resolving both the standard model and
new physics flavor puzzles at the Large Hadron Collider.
\end{abstract}

%11.30.Hv Flavor symmetries
%12.15.Ff Quark and lepton masses and mixing
%14.60.Pq Neutrino mass and mixing
%12.60.Jv Supersymmetric models
%13.85.-t Hadron-induced high- and super-high-energy interactions (> 10 GeV)

\maketitle

\section{The Flavor Puzzles}
\label{sec:intro}

The standard model (SM) of particle physics suffers from problems in
both its gauge and flavor sectors.  In the gauge sector, the Large
Hadron Collider (LHC) is expected to shed light on the hierarchy
problem, and the potential for discovering the Higgs boson and the
mechanism of electroweak symmetry breaking is largely responsible for
the keen anticipation for LHC data in the coming years.

In contrast, the LHC's prospects for explaining the {\em SM flavor
puzzle}, that is, the observed masses and mixings of the SM quarks and
leptons, are far less well-known.  Unlike the gauge sector puzzles,
the SM flavor puzzle is not necessarily connected to the weak scale.
In addition, data already stringently constrain most of the SM flavor
parameters, and it is not clear that more data will fix the problem.
As a case in point, the recent flood of data from neutrinos, far from
pointing the way to a compelling theory of flavor, has instead served
mainly to eliminate previous models, multiply the number of possible
explanations, and suggest that the origin of flavor is to be found at
very high energy scales.

What is often overlooked, however, is that new weak-scale physics may
shed light on the SM flavor puzzle in a very different way.  In many
well-motivated extensions of the SM, an understanding of the flavor
structure of new states can impose additional constraints on the same
set of theoretical parameters that govern the SM Yukawa couplings.
This is qualitatively different from the case of neutrinos, in that
new observables provide new constraints without introducing new
degrees of freedom.

A prominent example, and one we will consider in detail here, is the
case of supersymmetry.  For example, if SM flavor is determined by the
existence of an approximate horizontal symmetry (as in the
Froggatt-Nielsen mechanism~\cite{Froggatt:1978nt}),
then the SM particles and their
superpartners transform under the symmetry in the same
way.\footnote{If the flavor symmetry is an $R$-symmetry, the
transformation properties of the SM particles and their superpartners
are not the same, but still they are related.}  The same set of
horizontal charges will then be further constrained if existing flavor
changing neutral currents (FCNC) measurements are augmented by flavor
measurements at the LHC.  From this viewpoint, supersymmetry provides
a simple, representative example of new physics in which new particles
and the SM fermions behave identically under any underlying flavor
theory.  This feature is shared by other ideas for new physics,
including, for example, many models with extra dimensions, where the
flavor parameters of the SM fermions and their Kaluza-Klein excitations are
related.

Of course, new physics also brings with it the {\em new physics flavor
puzzle}: If there is new physics at the TeV scale, why does it not
contribute to FCNC processes at much higher rates than currently
observed~\cite{Nir:2007xn}? In the case of supersymmetry, for example,
if the supersymmetric flavor parameters are generic, then loop
diagrams involving gauginos and sfermions induce FCNC processes, such
as $K-\bar K$ mixing and $\mu\to e\gamma$, at levels that are orders
of magnitude above the experimentally allowed ranges. There are
essentially three mechanisms to suppress the supersymmetric
contributions to such processes:
\begin{itemize}
\item {\em Decoupling}. The sfermion mass scale can be very high. Such
  scenarios may be probed at the LHC through non-decoupling effects,
  for example, through the super-oblique
  parameters~\cite{Cheng:1997sq}.  However, as we are primarily
  interested here in the possibility of direct flavor measurements, we
  do not consider such scenarios further.
\item {\em Degeneracy}. The sfermion masses can be approximately
  degenerate, leading to GIM-like suppression. Such degeneracy could
  be the result of gauge-mediated supersymmetry breaking (GMSB) or of
  another type of flavor-blind mediation of supersymmetry breaking.
\item {\em Alignment}. The sfermion mixings, that is, the flavor-changing
  gaugino-sfermion-fermion couplings, can be suppressed.  Such
  alignment could be the result of an approximate horizontal symmetry.
\end{itemize}
Measurements of the supersymmetric flavor parameters --- the sparticle
masses and their flavor decomposition (mixing angles) --- will shed
light on the issue of how the new physics flavor puzzle is solved.  In
addition, as we will show below, there are many possible flavor models
that currently explain all available data, but which differ in their
resolution of the new physics flavor puzzle.  Determining how the new
physics flavor puzzle is resolved may therefore also play a key role
in leading us toward the correct solution to the SM flavor puzzle.

In this work, we focus on the slepton sector of supersymmetry. We
expect that the slepton sector is better suited for flavor
measurements at the LHC than the squark sector.  Many of the
theoretical issues that we raise also apply, however, to the squark
sector. Roughly speaking, the experimental upper bounds on the rates
of lepton flavor changing processes (\eg, $\mu\to e\gamma$) give
allowed ranges in the $(\Delta m^2_{ij}, K_{ij})$ planes. Here,
$\Delta m^2_{ij}$ stands for the mass-squared splitting between the
corresponding slepton generations, and $K_{ij}$ is the relevant mixing
angle. (Precise definitions are given below.) We argue that viable and
natural models can lead to many different points in the allowed range.
In the future, by combining the information from low and high energy
flavor measurements, we may be able to narrow the allowed range
considerably, select a specific supersymmetric model, and eventually
find the solution to both the new physics and SM flavor puzzles.

To demonstrate our point, we will analyze simple gauge-gravity
``hybrid'' models. (Other examples are possible, too.)  These are
minimal GMSB models with a high messenger scale, such that
gravity-mediated contributions cannot be neglected.  Since the GMSB
scalar masses are universal, the gravity-mediated contributions
dictate the mixings, even if their overall size is quite small, and so
we will invoke horizontal symmetries to adequately suppress the
mixings. The models we present are consistent with all known lepton
masses and mixings, and they satisfy all FCNC and rare decay
constraints. At the same time, they span a wide range of predictions
for slepton flavor violation, and are amenable to many LHC flavor
studies, providing a useful starting point for our purposes.

The flavor problems associated with supersymmetry, and more generally
with TeV-scale new physics, have led to the spreading belief in the
flavor physics community that minimal flavor violation (MFV) is
perhaps inescapable. The models we present demonstrate that this is
far from being true. One of the general conclusions of this work is
that the question of whether new physics is MFV or non-MFV is open,
and it is therefore of great interest to determine the LHC's potential
for distinguishing between models that exhibit MFV and models that do
not.\footnote{The LHC has the potential to test MFV also in the
context of other extensions of the SM~\cite{Grossman:2007bd}.}

\section{Phenomenological Constraints}
\label{sec:phecon}

Rare flavor-changing charged lepton decays constrain a combination of
supersymmetric parameters. It is a reasonable approximation, for our
purposes, to think of the constrained quantities as
\begin{equation}
\delta^M_{ij}\sim \frac{\Delta\tilde m^{2}_{Mji}}{\tilde
  m^{2}_M}K_{ij}^M \ ,
\end{equation}
where $M=L,R$ specifies left-handed sleptons $\tilde E_L$ or
right-handed sleptons $\tilde E_R$, $i,j=1,2,3$ are generation
indices, and
\begin{eqnarray}
\tilde m_M&=&(m_{\tilde E_{Mi}}+m_{\tilde E_{Mj}})/2 \ , \no\\
\Delta\tilde m^{2}_{Mji}&=&m^2_{\tilde E_{Mj}}-m^2_{\tilde E_{Mi}} \ .
\end{eqnarray}
The matrix $K^M$ is the mixing matrix of electroweak gaugino
couplings. In other words, $g_\alpha K^M_{ij}$ is the coupling
strength of the $\lambda_\alpha$ gaugino ($\alpha=1,2$ for the Bino,
Wino) to the lepton $E_{Mi}$ and the slepton $\tilde E_{Mj}$. We
ignore here the constraints on the LR block in the slepton
mass-squared matrix, which are, however, satisfied in our models.

In Ref.~\cite{Ciuchini:2007ha}, the experimental bounds\footnote{The
upper bound on $\tau\to\mu\gamma$ has since been
improved~\cite{Hayasaka:2007vc} to $B( \tau \to \mu \gamma) \leq 4.5
\times 10^{-8}$.}
\begin{eqnarray}
B(\mu\to e\gamma)&\leq&1.2\times10^{-11} , \quad
B(\tau\to e\gamma)\leq1.1\times10^{-7} , \quad
B(\tau\to\mu\gamma)\leq6.8\times10^{-8} ,
\end{eqnarray}
were used to derive the following constraints:
\begin{eqnarray}\label{delbou}
\delta^{L}_{12}&\leq&6\times10^{-4} \ , \quad
\delta^{R}_{12}\leq0.09 \ , \quad
\delta^{M}_{13}\leq0.15 \ , \quad
\delta^{M}_{23}\leq0.12 \ .
\end{eqnarray}
These results assumed a universal scalar mass $m_0 < 380~\gev$, a
unified gaugino mass $M_{1/2} < 160~\gev$, and $5 < \tan\beta < 15$.
We are interested in a more generic framework that does not assume
{\em a priori} universality or unification. Nevertheless, we will
require that our models satisfy the constraints of \eqref{delbou},
because the numerical details are not significant for our purposes. In
particular, they can be modified in a straightforward way to meet
stronger (or milder) constraints.  Our models also reproduce the
present ranges of the leptonic mixing angles:
\begin{equation}\label{mixl}
\sin^2\theta_{12}=0.31 \pm 0.02 \ , \quad
\sin^2\theta_{13}=0^{+0.008}_{-0.0} \ , \quad
\sin^2\theta_{23}=0.47 \pm 0.07 \ .
\end{equation}

Our models employ a small symmetry breaking parameter, $\lambda \sim
0.1-0.2$. The various physical parameters are suppressed by powers of
$\lambda$, with unknown ${\cal O}(1)$ coefficients. We thus interpret
the constraints of \eqref{delbou}, the measurements of \eqref{mixl},
and the information on lepton masses in terms of $\lambda$ as follows:
\begin{eqnarray}
\delta^{L}_{12} &\alt& \lambda^4 \ , \quad
\delta^{R}_{12} \alt \lambda^2 \ , \quad
\delta^{M}_{13} \alt \lambda \ , \quad
\delta^{M}_{23} \alt \lambda \ , \quad
\sin\theta_{ij} \sim 1 \ , \no \\
m_1/m_2 &\sim& 1 \ , \quad m_2/m_3 \sim 1 \ , \quad
m_e/m_\mu \sim \lambda^2 \ , \quad m_\mu/m_\tau \sim \lambda^2 \ .
\end{eqnarray}
Note that we assume here that $\sin\theta_{13}$ is accidentally,
rather than parametrically, suppressed. If the experimental upper
bound on $\sin\theta_{13}$ improves significantly, the models would
have to be modified.

\section{Minimal Flavor Violation}
\label{sec:mfv}

The supersymmetric flavor puzzle is solved if the mediation of
supersymmetry breaking obeys the principle of
MFV~\cite{D'Ambrosio:2002ex}. In MFV models, in the absence of the SM
Yukawa couplings (possibly extended to allow for neutrino masses), the
leptonic sector has a global SU(3)$_L \times$ SU(3)$_E$ symmetry.  The
SU(3)$_L$ acts on the three SU(2)-doublet lepton supermultiplets, and
the SU(3)$_E$ acts on the three charged SU(2)-singlet lepton
supermultiplets. The symmetry is broken by the charged lepton Yukawa
couplings, which constitute spurions transforming as $(3,\bar3)$ under
the global symmetry. There could be additional spurions related to
neutrino masses~\cite{Cirigliano:2005ck}.  For the sake of simplicity,
we assume that these additional spurions can be neglected. Indeed,
they are expected to be negligible whenever the scale of mediation of
supersymmetry-breaking is much lower than the scale of lepton number
breaking (the seesaw scale). In any case, the possible effects of such
spurions can be included in a straightforward way. Our conclusions for
right-handed sleptons are hardly affected by such spurions.

Within MFV, we can choose to work in a basis where the $Y_E$ spurion
is diagonal,
\begin{equation}
Y_E=\text{diag}\, (y_e,y_\mu,y_\tau) \ .
\end{equation}
If the spurions related to the neutrino masses are negligible, the soft
slepton mass-squared terms have the form
\begin{equation}\label{mslr}
M^2_{\tilde E_M}\sim\tilde m^2_M\pmatrix{
  1+a_M y_e^2 & 0 & 0 \cr 0 & 1+a_My_\mu^2 & 0 \cr
  0& 0 & 1+a_My_\tau^2\cr},
\end{equation}
where the dimensionless coefficients $a_M$ are $\alt 1$. The spectrum
and flavor decomposition of sleptons in MFV models therefore have the
following properties:
\begin{enumerate}
\item The spectrum has three-fold degeneracy to a good approximation.
  The fractional mass splitting of the third generation is of order
  $y_\tau^2$.  For $\tan\beta\sim m_t/m_b$, $\tilde\tau_{L,R}$ is
  split from $\tilde e_{L,R}$ and $\tilde\mu_{L,R}$ by roughly
  10\%. For smaller $\tan\beta$, the splitting scales down as
  $\tan^2\beta$.
\item The first two generations are degenerate to an excellent
  approximation, as their fractional mass splitting is of order
  $y_\mu^2$.
\item There is no mixing.
\end{enumerate}

\section{Non-Minimal Flavor Violation}
\label{sec:nonmfv}

In this section we argue that there is much room for flavor physics
that is far from MFV. To do so, we present explicit models that are
both natural (\ie, small couplings are related to approximate
symmetries) and viable, and yet violate some or all of the MFV
predictions listed above in a significant way. These models are based
on balancing two types of contributions to the slepton mass matrices,
in the spirit of Ref.~\cite{Feng:2007sx}: a gauge-mediated
contribution, which is MFV, and a gravity-mediated contribution that
is non-MFV. We assume, however, that the gravity-mediated contribution
is subject to an approximate horizontal symmetry and can thus exhibit
approximate alignment~\cite{Nir:1993mx}.  Thus, the supersymmetric
flavor problem is solved by a combination of degeneracy and alignment.
We construct models that saturate the upper bound on the
$\delta^L_{12}$ parameter, but the models can be easily modified to
give contributions that are well below the present bound.

We further require that the horizontal symmetry accounts for the
observed flavor features of leptons. Specifically, we require that the
symmetry gives anarchical neutrino mass matrices (with neither
hierarchy nor degeneracy in the neutrino masses), ${\cal O}(1)$
leptonic mixings, and hierarchical charged lepton masses. Again, our
models can be modified in a straightforward way to provide parametric
suppressions that are different from these choices. Thus, both the
approximate alignment of the sleptons and the structure of the lepton
masses and mixings are dictated by the same symmetry and the same
horizontal charges. As explained in \secref{intro}, this is exactly
the sort of scenario in which experimental information on slepton
mixing will contribute to our understanding of the SM flavor puzzle.

Let us be more specific now. Most of our models are high-scale GMSB
models such that the soft masses are dominated by the GMSB
contributions, with a smaller component of gravity-mediated masses.
It is convenient to parameterize the ratio between the
gravity-mediated and the gauge-mediated contribution by $x$:
\begin{eqnarray}\label{param}
M^2_{\tilde E_L}&=&\tilde m^2_L{\bf 1}+m_Em_E^\dagger+x\tilde m^2_L
X_L\ , \label{xl}\\
M^2_{\tilde E_R}&=&\tilde m^2_R{\bf 1}+m_E^\dagger m_E+x\tilde m^2_R
X_R\ .
\end{eqnarray}
The GMSB contributions are universal, but the gravity-mediated
contributions result in potentially large mixings.  We invoke
horizontal (Abelian) symmetries to suppress these mixings to
acceptable levels, in the spirit of alignment
models~\cite{Nir:1993mx}.  The SM matter fields are charged under the
horizontal symmetry, and the symmetry is assumed to be spontaneously
broken, with the breaking parameterized by one or more spurion fields
whose vacuum expectation values are smaller than 1.  Since the full
Lagrangian must respect the horizontal symmetry, each term in the
Lagrangian must involve an appropriate power of the spurion(s).

We assume that supersymmetry breaking is dominated by a single
$F$-term, which contributes both to gauge mediation and gravity
mediation. The GMSB contributions to the left-handed slepton
masses-squared are universal ($\tilde m_L^2{\bf 1}$), with
\begin{equation}\label{gmsb}
\tilde{m}_L^2 \sim N_m \left(\frac{\alpha_2}{\pi}\right)^2
\left(\frac{F}{M_m}\right)^2\ ,
\end{equation}
where $N_m$ is the number of $5+\bar5$ messenger pairs, and $M_m$ is
the messenger scale. The gravity-mediated contributions arise from
the K\"ahler terms
\begin{equation}\label{gm}
{X_L}_{ij} \frac{S^\dagger S}{M_P^2}\, L_i^\dagger L_j \ ,
\end{equation}
where $S$ is the SM-singlet whose $F$-term triggers supersymmetry
breaking, and $M_P \simeq 2.4 \times 10^{18}~\gev$ is the reduced
Planck scale.  The coefficients ${X_L}_{ij}$ are given, up to ${\cal
O}(1)$ numbers, by the appropriate power of the horizontal symmetry
spurion. This power is determined by the charges of $L_i$ and $L_j$,
so that off-diagonal elements in the mass matrices are generically
suppressed compared to the diagonal elements.  The terms of \eqref{gm}
give rise to the soft masses
\begin{equation}
{X_L}_{ij}\, \left(\frac{F}{M_P}\right)^2 \ .
\end{equation}
We can estimate the messenger scale $M_m$ as a function of
$x$,\footnote{Strictly speaking, we would get a somewhat different
estimate if we used $\tilde{m}_R^2$ instead of $\tilde{m}_L^2$, since
the former involves $\alpha_1$ and the latter involves $\alpha_2$,
with different numerical coefficients. However, at the high messenger
scales we are considering, these two couplings are not very different,
and we are neglecting ${\cal O}(1)$ coefficients throughout
anyway. Furthermore, the only part of our analysis that is sensitive
to the messenger scale is the NLSP slepton lifetime.  Regardless of
these ${\cal O}(1)$ differences, the NLSP will always decay outside
the detector in our models.  }
\begin{equation}
M_m\sim \sqrt{N_m \cdot x}\, \frac{\alpha_2}\pi\, M_P \ ,
\end{equation}

Examining \eqref{param}, we note that it can lead to degeneracy,
$\Delta \tilde m^2_{Mji}/\tilde m^2_M\sim x$ for small $x$, and to alignment,
$K^M_{ij}\sim \text{max}[(X_M)_{ij},(V_M^E)_{ij}]$, where $V_L^E M_E
V_R^{E\dagger}=\text{diag}(m_e,m_\mu,m_\tau)$. We learn that
\begin{equation}
\delta^M_{ij} \sim x \times \text{max} \left[ (X_M)_{ij},(V_M^E)_{ij}
\right] \ .
\end{equation}
Thus, if the horizontal symmetry produces strong alignment (small
mixing angles), then the degeneracy can be mild ($x\sim1$), but if the
supersymmetric mixing angles are large (similar to the leptonic mixing
angles $\theta_{12}$ and $\theta_{23}$), then the degeneracy must be
strong ($x \ll 1$). We present four different models that demonstrate
the various possibilities, from strong alignment/no degeneracy to
large mixing/small splittings.

We assume a horizontal U(1) $\times$ U(1) symmetry, where each of the
U(1)'s is broken by a spurion of corresponding charge $-1$ and size
$\lambda \sim 0.2$. (Our results would remain the same if the first
U(1) were broken by two spurions of charges $\pm 1$.) For each model,
we give the horizontal charges of the left-handed lepton doublet $L_i$
and antilepton singlet $\bar E_i$ supermultiplets (setting the Higgs
charges to zero), the lepton mass matrices and the gravity-mediated
contribution to the slepton mass-squared matrices (omitting
coefficients of order one in each entry), the parametric suppression
(\ie, the $\lambda$-dependence) of the resulting mixing angles, and
the maximum allowed value of $x$, which gives the level of degeneracy.

%%%%%%%%%%%%%%
\subsection{Small mixing, no degeneracy}
The first model has pure gravity mediation, supplemented
by the horizontal symmetry. More generally, there can be small
gauge-mediated contributions, but they are at most comparable to
the gravity-mediated ones.

The horizontal U(1) $\times$ U(1) charges are
\begin{equation}
L_1(4,0),\ L_2(2,2),\ L_3(0,4) ;\quad \bar E_1(1,0),\ \bar E_2(1,-2),\
\bar E_3(0,-3) \ .
\end{equation}
The resulting lepton mass matrices have the following
structure:\footnote{The zeros in these mass matrices follow from
holomorphy. The vanishing entries would require powers of
$\lambda^\dagger$ to form U(1) $\times$ U(1)-invariant combinations
but, since the superpotential is holomorphic, $\lambda^\dagger$ cannot
appear. }
\begin{equation}
M_\nu\sim\frac{\langle\phi_u\rangle^2}{M}\lambda^8\pmatrix{
  1&1&1\cr 1&1&1\cr 1&1&1\cr} \ , \quad
M_E\sim\langle\phi_d\rangle\lambda\pmatrix{\lambda^4 & 0 & 0\cr
  \lambda^4 & \lambda^2 & 0 \cr \lambda^4 & \lambda^2 & 1 \cr}.
\end{equation}
The $X_M$ matrices have the following structure:
\begin{equation}
X_L \sim \pmatrix{
  1 & \lambda^4 & \lambda^8 \cr \lambda^4 & 1 & \lambda^4 \cr
  \lambda^8 & \lambda^4 & 1 \cr} \ , \quad
X_R\sim \pmatrix{
  1 & \lambda^2 & \lambda^4 \cr \lambda^2 & 1 & \lambda^2 \cr
  \lambda^4 & \lambda^2 & 1 \cr}.
\end{equation}
The parametric suppression of the mixing angles is given by
\begin{equation}
K^L_{12}\sim\lambda^4,\ \ K^L_{13}\sim\lambda^8,\ \
K^L_{23}\sim\lambda^4;\ \ \
K^R_{12}\sim\lambda^2,\ \ K^R_{13}\sim\lambda^4,\ \
K^R_{23}\sim\lambda^2 \ .
\end{equation}
There is no degeneracy:
\begin{equation}
x\agt 1 \ ,
\end{equation}
as is the case for
\begin{equation}
M_m\agt  0.1 M_P \ .
\end{equation}
Given this high messenger scale, and the fact that the gravity- and
gauge-mediated contributions are comparable, we will think of this
model as a pure gravity-mediation model with no GMSB contribution.

The flavor suppression in this model comes entirely from the
smallness of the supersymmetric mixing angles. In other words,
the alignment of the charged slepton mass eigenstates with the
charged lepton mass eigenstates is precise enough to satisfy all
phenomenological constraints. No degeneracy is required.

%%%%%%%%%%%%%%
\subsection{Large 2-3 mixing, small 1-2 mixing, ${\cal O}(0.1)$ degeneracy}
This model is a gauge-gravity hybrid model.
The suppression in the $2-3$ sector is provided by degeneracy, while
in the $1-2$ sector it comes mainly from alignment.

The horizontal U(1) $\times$ U(1) charges are
\begin{equation}
L_1(2,0),\ L_2(0,2),\ L_3(0,2);\ \ \ \bar E_1(2,1),\ \bar E_2(2,-1),\
\bar E_3(0,-1).
\end{equation}
The resulting lepton mass matrices have the following structure:
\begin{equation}
M_\nu\sim\frac{\langle\phi_u\rangle^2}{M}\lambda^4\pmatrix{
  1&1&1\cr 1&1&1\cr 1&1&1\cr} \ , \quad
M_E\sim\langle\phi_d\rangle\lambda\pmatrix{\lambda^4 & 0 & 0\cr
  \lambda^4 & \lambda^2 & 1 \cr \lambda^4 & \lambda^2 & 1 \cr}.
\end{equation}
The $X_M$ matrices have the following structure:
\begin{equation}
X_L\sim\pmatrix{
  1 & \lambda^4 & \lambda^4 \cr \lambda^4 & 1 & 1 \cr
  \lambda^4 & 1 & 1 \cr} \ , \quad
X_R\sim\pmatrix{
  1 & \lambda^2 & \lambda^4 \cr \lambda^2 & 1 & \lambda^2 \cr
  \lambda^4 & \lambda^2 & 1 \cr}.
\end{equation}
The parametric suppression of the mixing angles is given by
\begin{equation}
K^L_{12}\sim\lambda^4,\ \ K^L_{13}\sim\lambda^4,\ \
K^L_{23}\sim1;\ \ \
K^R_{12}\sim\lambda^2,\ \ K^R_{13}\sim\lambda^4,\ \
K^R_{23}\sim\lambda^2.
\end{equation}
The level of degeneracy required for satisfying the
$\delta^L_{23}$ constraint is rather mild:
\begin{equation}
x\sim0.1 \ ,
\end{equation}
leading to
\begin{equation}
M_m\sim 10^{-3} M_P \ .
\end{equation}

%%%%%%%%%%%%%%
\subsection{Large 2-3 mixing, mildly small 1-2 mixing, ${\cal O}(0.02)$
degeneracy}
This is another gauge-gravity hybrid model.  The horizontal U(1)
$\times$ U(1) charges are
\begin{equation}
L_1(1,0),\ L_2(0,1),\ L_3(0,1);\ \ \ \bar E_1(2,1),\ \bar E_2(2,-1),\
\bar E_3(0,-1).
\end{equation}
The resulting lepton mass matrices have the following structure:
\begin{equation}
M_\nu\sim\frac{\langle\phi_u\rangle^2}{M}\lambda^2\pmatrix{
  1&1&1\cr 1&1&1\cr 1&1&1\cr} \ , \quad
M_E\sim\langle\phi_d\rangle\pmatrix{\lambda^4 & 0 & 0\cr
  \lambda^4 & \lambda^2 & 1 \cr \lambda^4 & \lambda^2 & 1 \cr}.
\end{equation}
The $X_M$ matrices have the following structure:
\begin{equation}
X_L\sim\pmatrix{
  1 & \lambda^2 & \lambda^2 \cr \lambda^2 & 1 & 1 \cr
  \lambda^2 & 1 & 1 \cr} \ , \quad
X_R\sim\pmatrix{
  1 & \lambda^2 & \lambda^4 \cr \lambda^2 & 1 & \lambda^2 \cr
  \lambda^4 & \lambda^2 & 1 \cr}.
\end{equation}
The parametric suppression of the mixing angles is given by
\begin{equation}
K^L_{12}\sim\lambda^2,\ \ K^L_{13}\sim\lambda^2,\ \
K^L_{23}\sim1;\ \ \
K^R_{12}\sim\lambda^2,\ \ K^R_{13}\sim\lambda^4,\ \
K^R_{23}\sim\lambda^2.
\end{equation}
The required level of degeneracy is dictated by the $\delta^L_{12}$
constraint:
\begin{equation}
x\sim0.02 \ ,
\end{equation}
leading to
\begin{equation}
M_m\sim 10^{-4}M_P \ .
\end{equation}

%%%%%%%%%
\subsection{Strong degeneracy, large mixing}
We can also take a single horizontal U(1) and assign all left-handed
lepton doublets the same horizontal charge:
\begin{equation}
L_1(2),\ L_2(2),\ L_3(2);\ \ \ \bar E_1(4),\ \bar E_2(2),\
\bar E_3(0).
\end{equation}
The resulting lepton mass matrices have the following structure:
\begin{equation}
M_\nu\sim\frac{\langle\phi_u\rangle^2}{M}\lambda^4\pmatrix{
  1&1&1\cr 1&1&1\cr 1&1&1\cr} \ , \quad
M_E\sim\langle\phi_d\rangle\lambda^2\pmatrix{\lambda^4 & \lambda^2 & 1\cr
  \lambda^4 & \lambda^2 & 1 \cr \lambda^4 & \lambda^2 & 1 \cr}.
\end{equation}
The $X_L$ matrix is anarchical:
\begin{equation}
X_L\sim\pmatrix{
  1 & 1 & 1 \cr 1 & 1 & 1 \cr
  1 & 1 & 1 \cr} \ , \quad
X_R\sim\pmatrix{
  1 & \lambda^2 & \lambda^4 \cr \lambda^2 & 1 & \lambda^2 \cr
  \lambda^4 & \lambda^2 & 1 \cr}.
\end{equation}
The parametric suppression of the mixing angles is given by
\begin{equation}
K^L_{12}\sim1,\ \ K^L_{13}\sim1,\ \
K^L_{23}\sim1;\ \ \
K^R_{12}\sim\lambda^2,\ \ K^R_{13}\sim\lambda^4,\ \
K^R_{23}\sim\lambda^2.
\end{equation}
The required suppression must come entirely from degeneracy:
\begin{equation}
x\sim0.001 \ ,
\end{equation}
leading to
\begin{equation}
M_m\sim 10^{-4}M_P \ .
\end{equation}

%%%%%%%%%%%%%%%%%%
\subsection{Features of the $\tilde E_R$ sector}
A summary of the spectra and the mixings in the $\tilde E_R$ sector in
MFV and in our non-MFV models is presented in Table \ref{tab:spemixr}.
We conclude:
\begin{enumerate}
\item Measurements of upper bounds on mass splitting at any level will
  be informative.
\item If mass splitting is established and $\tan\beta$ is not large,
  that will clearly signal non-MFV.
\item Mixing effects are small, ${\cal O}(\lambda^2)$, with or without
 MFV.  If one could be sensitive to $\tilde e_R-\tilde\mu_R$ mixing
 ($\tilde \mu_R-\tilde\tau_R$ mixing) of order a (few) percent, that
 would clearly distinguish MFV from non-MFV.
 \end{enumerate}

\begin{table}[t]
\caption{Spectrum and mixing of charged slepton singlets.
The vectors below the mass eigenstate $\tilde{E}_{Ri}$ give its
flavor decomposition, namely the $(\tilde e_R,\tilde\mu_R,\tilde\tau_R)$
components of $\tilde E_{Ri}$.}
\label{tab:spemixr}
\begin{center}
\begin{tabular}{c|cc|ccc} \hline\hline
\rule{0pt}{1.2em}%
%\label{tab:bqqq}
%\settabs 5 \columns
Model & $\Delta\tilde m_{R21}/\tilde m_R$ & $\Delta\tilde m_{R32}/\tilde
 m_R$ & $\tilde E_{R1}$ &  $\tilde E_{R2}$ & $\tilde E_{R3}$ \cr \hline\hline
MFV & $10^{-6}\tan^2\beta$ & $10^{-4}\tan^2\beta$ & $(1,0,0)$ &
 $(0,1,0)$ & $(0,0,1)$  \cr
A, B, C, D & $\alt 1$ & $\alt 1$ & $(1,0.01,0.001)$ & $(0.01,1,0.1)$ &
$(0.001,0.1,1)$  \cr
 \hline\hline
\end{tabular}
\end{center}
\end{table}

We stress that while we did not make an effort to vary the level of
mixing in the $\tilde E_R$ sector in our models, we expect this mixing
to be small model independently. The reason is that in models of
Abelian horizontal symmetries, we have a generic upper bound on the
mixing~\cite{Grossman:1995hk,Feng:1999wt}:
\begin{equation}
K_{ij}^R\alt  (m_{\ell_i}/m_{\ell_j})/|U_{ij}| \ ,
\end{equation}
where $U$ is the lepton mixing matrix.
Given our assumptions that the lepton mixings have no
$\lambda$-suppression while the charged lepton masses have a
$\lambda^2$-hierarchy, then all our non-MFV models satisfy the naive
upper bound on the slepton mixing.  In particular, $K_{ij}$ cannot be
order 1.

%%%%%%%%%%%%%%%%%%
\subsection{Features of the $\tilde E_L$ Sector}

We summarize our results in Table \ref{tab:spemix}. We give the mass
splittings and the flavor decompositions of the mass eigenstates for
the MFV framework and for each of our four non-MFV examples.  We draw
the following conclusions:
\begin{enumerate}
\item Mass measurements with an accuracy of ${\cal O}(0.1-0.001)$
(which, for 200 GeV sleptons, translates into mass resolutions of
$20-0.2$ GeV) will provide valuable information.
\item Of course, it will be possible to probe the flavor decomposition
  only if the mass eigenstates can be separated. In this case,
  \begin{itemize}
\item It would be helpful to learn whether the mass eigenstate that
  decays dominantly into tau-leptons (muons) decays to muons
  (tau-leptons) at a comparable rate or a much smaller rate.
\item It would be informative to measure (or put an upper bound on)
  the $\mu$ and/or $\tau$ branching ratio of the mass eigenstate that
  decays dominantly to electrons.
\end{itemize}\end{enumerate}

\begin{table}[t]
\caption{Spectrum and mixing of slepton doublets.  The vectors below
the mass eigenstate $\tilde{E}_{Li}$ give its flavor decomposition.
For the starred quantities, the entry is the maximum of the written
value and the MFV value.}
\label{tab:spemix}
\begin{center}
\begin{tabular}{c|cc|ccc} \hline\hline
\rule{0pt}{1.2em}%
%\label{tab:bqqq}
%\settabs 5 \columns
Model & $\Delta\tilde m_{L21}/\tilde m_L$ & $\Delta\tilde m_{L32}/\tilde
 m_L$ & $\tilde E_{L1}$ &  $\tilde E_{L2}$ & $\tilde E_{L3}$ \cr \hline\hline
MFV & $10^{-6} \tan^2\beta$ & $10^{-4} \tan^2\beta$ & $(1,0,0)$ &
 $(0,1,0)$ & $(0,0,1)$  \cr
A & $1$ & $1$ & $(1,10^{-3},10^{-6})$ & $(10^{-3},1,10^{-3})$ &
$(10^{-6},10^{-3},1)$  \cr
B & $0.1$ & $0.1^*$ & $(1,10^{-3},10^{-3})$ & $(10^{-3},1,1)$ &
$(10^{-3},1,1)$  \cr
C & $0.02$ & $0.02^*$ & $(1,0.04,0.04)$ & $(0.04,1,1)$ &
$(0.04,1,1)$  \cr
D & $0.001$ & $0.001^*$ & $(1,1,1)$ & $(1,1,1)$ &
$(1,1,1)$  \cr
 \hline\hline
\end{tabular}
\end{center}
\end{table}

\section{The LSP and NLSP}
\label{sec:lsp}

Although our models differ in their flavor structures, they share the
feature of a gravitino LSP.  In addition, for moderate to large $N_m$,
the NLSP is a charged slepton~\cite{Feng:1997zr}. The reason for that
is that, in our framework,
\begin{equation}
m_{3/2}\sim \sqrt{x} \, \tilde{m}_{\text{slepton}} \ .
\end{equation}
Thus, the only possible exception is model A, where $x \sim 1$, and
the slepton and gravitino masses are consequently comparable. But even
in this model it is possible that the gravitino is ``accidentally''
lighter than the charged sleptons.

The decay width for slepton decay to gravitino is independent of the
slepton's flavor and chirality composition, and is given by
\begin{equation}
\Gamma ( \tilde{E}_M \to l \tilde{G} ) = \frac{1}{48 \pi M_P^2}
\frac{m_{\tilde{E}_M}^5}{m_{\tilde{G}}^2}
\left[ 1 - \frac{m_{\tilde{G}}^2}{m_{\tilde{E}_M}^2} \right]^4 \ .
\end{equation}
For $m_{\tilde{G}} \ll m_{\tilde{E}_M}$, the slepton lifetime is
\begin{equation}
\tau \simeq 16~\text{hours} \
\left[ \frac{m_{\tilde{G}}}\gev \right]^2 \
\left[ \frac{100~\gev}{m_{\tilde{E}_M}} \right]^5 \ .
\end{equation}
For all of the models considered here, then, the slepton is
effectively stable for collider experiments.  Supersymmetric events
are fully reconstructible, and, at least in principle, the final state
jets and leptons can be combined to form the intermediate
supersymmetric particles all the way up the cascade decay chains.

This scenario differs from the usual supersymmetric scenario, in which
supersymmetric events are characterized by two missing neutralinos.
Slepton flavor violation at future colliders in missing energy
scenarios has been the subject of many studies.  (See, \eg,
Refs.~\cite{ArkaniHamed:1996au,ArkaniHamed:1997km,Bityukov:1997ck,%
Agashe:1999bm,Hinchliffe:2000np,Hisano:2002iy}.)  These results will
be improved markedly in the scenario discussed here, where
supersymmetric events are fully reconstructible.  In particular, final
state leptons may be identified as originating from interactions with
left- or right-handed sleptons.  Thus, the LHC may be able to
determine the slepton-lepton-gaugino mixing angles in both left- and
right-handed sectors independently, providing extremely incisive tests
of all of the flavor models presented above.

In addition, for the high-scale gauge mediation models discussed here,
the extremely long lifetime of the slepton implies that it may be
possible to trap and collect sleptons and observe their
decays~\cite{Feng:2004yi,Hamaguchi:2004df,De Roeck:2005bw}.  Collider
measurements of the slepton mass and a determination of the slepton
lifetime determines the gravitino mass.  (In fact, a measurement of
the lepton energy in slepton decays provides another measurement of
the gravitino mass, or alternatively, a check that the outgoing
particle couples with gravitational
strength~\cite{Buchmuller:2004rq,Feng:2004gn}.)  The gravitino mass is
a measure of the $F$-term relevant for gravity mediation, and
comparison with the $F$-term relevant for gauge-mediation will provide
useful insights into supersymmetry breaking.

Of course, the observation of NLSP decay also provides flavor
information. Comparison of the decay rate of the NLSP to $e
\tilde{G}$, $\mu \tilde{G}$, and $\tau \tilde{G}$ provides a direct
measurement of the flavor composition of the NLSP.  Depending on the
number of NLSP decays observed, this may provide precise constraints
on slepton flavor violation~\cite{Hamaguchi:2004ne}, which will
supplement the other flavor information derived directly from collider
experiments.

\section{Slepton Flavor at the LHC}
\label{sec:exp}

We have demonstrated that measurements of (or bounds on) mass
splittings and mixings in the slepton sector will be very valuable for
our understanding
of flavor and of supersymmetry breaking. Can these
goals be achieved at the LHC?
We postpone a detailed study of this question to future
work~\cite{inprogress}, and confine ourselves here to brief comments on the
opportunities and limitations of the ATLAS and CMS experiments for LFV
studies.

There have been several suggestions in
the literature for signal channels at future colliders in which to
look for lepton flavor violation~\cite{ArkaniHamed:1996au,%
ArkaniHamed:1997km,Bityukov:1997ck,Agashe:1999bm,Bartl:2005yy,%
Deppisch:2007rm}, and a few works that study whether such signals will
actually be measurable at the
LHC~\cite{Hinchliffe:2000np,Hisano:2002iy,Assamagan:2002kf}.
There are, however, a number of reasons to revisit this question now.
The ATLAS and CMS collaborations have recently made great progress by
(1) tying their detector simulations more closely to the as-built
geometries and material budgets, (2) replacing back-of-the envelope
estimates of reconstruction performance (such as vertexing resolutions
and charged lepton and jet resolutions and efficiencies) with models
based on full simulations and algorithms approaching those that will
be used for the final detector, (3) modeling the trigger response, and
(4) generating a wider range and greater number of background samples
than were available a few years ago.  With regard to background
samples, the latest next-to-leading-order parton shower-matching event
generators~\cite{Frixione:2002ik,Frixione:2003ei,Mangano:2002ea} have
yielded more accurate production of multi-parton SM processes.

Such advances are of great relevance for LFV studies.  For example,
before particle misidentification is taken into account, the
multi-lepton signatures generic in many LFV studies will have low SM
backgrounds.  However, in reality, jets misidentified as leptons may
create significant backgrounds, given the gigantic QCD cross section
at the LHC.  Proper modeling of jet misidentification is therefore
highly relevant for all LFV studies, and particularly those that use
tau leptons.

Most previous studies of LFV have considered missing energy signals.
As noted in \secref{lsp}, however, in the hybrid gauge-gravity
mediation scenarios considered here, long-lived charged sleptons are
at least as likely, and it is possible, at least in principle, to
reconstruct the four-momenta and invariant masses of all particles in
each event on an event-by-event
basis~\cite{Ellis:2006vu,Polesello:683824,Ambrosanio:2000ik,%
Zalewski:2007up}. Of course, it is essential to determine whether
pre-selection cuts may be devised to isolate a relatively pure sample
of such events, and to assess the precision with which masses and mass
splittings may be measured in such scenarios.  Such studies will
require adequate modeling of backgrounds and lepton and jet
reconstruction, as well as trigger efficiencies for slow metastable
sleptons~\cite{Bressler:2007gk}.

Finally, it is also of great interest to investigate the experimental
sensitivity to mass splittings in situations where flavor mixing is
small, and when the final supersymmetric particles are not seen by the
detector, as may be the case in our model A~\cite{inprogress}.
Since the direct reconstruction of the previous paragraph is no
  longer possible, it may be necessary to use one or more of the
  indirect techniques for placing constraints on supersymmetric
  events with incomplete information which have been suggested in the
  past~\cite{Paige:1996nx,Hinchliffe:1998ys,Bachacou:1999zb,
  Allanach:2000kt,Nojiri:2003tu,Allanach:2004ub,
  Gjelsten:2005aw,Miller:2005zp,Lester:2006cf}.
We plan to check whether they are applicable to our
framework. For our purposes, the principal idea is to plot invariant
mass spectra of dilepton pairs, and look for an upper limit (a
cut-off) in their distributions~\cite{Paige:1996nx}. For example, a
difference in the end-point positions of the $m_{e^+ e^-}$ and
$m_{\mu^+ \mu^-}$ distributions constrains the mass-squared
differences of the selectron and the smuon, even though the absolute
value of either mass would be almost unconstrained.

\section{Discussion}
\label{sec:discussion}

It is convenient to think about the issue of supersymmetric lepton
flavor in the $(\Delta m^2_{ij}, K_{ij})$ plane, \ie, the mass-squared
splitting between slepton generations {\em vs.} the mixing among them.

>From the experimental point of view, we can probe this plane in at
least two independent ways. Upper bounds on rare charged lepton
decays, such as $\mu\to e\gamma$, give an upper bound on the
combination $\Delta m^2_{ij}\times K_{ij}$, which corresponds to a
curve in the plane. The lower left region below this curve is
allowed. If future experiments actually measure these rates, they will
confine models to the corresponding curve.

Direct observations of sleptons at the ATLAS and CMS experiments can
provide upper bounds on either or both of the mass splitting and
mixing. Under favorable circumstances, both can be measured. It is
thus not inconceivable that some combination of lepton flavor
precision measurements and direct measurements at the LHC will
eventually allow only a small region in the $(\Delta m^2_{ij},
K_{ij})$ plane.

>From the theoretical point of view, the supersymmetric ``model space''
is rich enough that it can (naturally!) lead to almost any point in
the plane. GMSB contributions to the soft breaking terms lead to
degeneracies that can be as small as the corresponding lepton
mass-squared (\eg, $m_\mu^2$ between sleptons of the first and second
generations). The larger the gravity-mediated contribution, the larger
the splitting would be, and it can reach order one if these
contributions are comparable to or larger than the gauge-mediated
ones.

Horizontal symmetries can lead to alignment. Depending on the symmetry
and on the horizontal charges, the mixing in the left-handed sector
can be as large as the corresponding lepton mixing, or much
smaller. The mixing in the right-handed sector can be as large as the
ratio between the lepton mass-ratio and the mixing angle, \ie,
$K^R_{ij} \alt (m_{\ell_i}/m_{\ell_j})/U_{ij}$
(here $U$ is the mixing matrix in the lepton sector).
Either mixing can also be much smaller than these upper bounds.

Thus, the ratio between gauge- and gravity-mediated contributions
allows us to move along the $\Delta m^2_{ij}$ axis, while the
horizontal charge assignments allow us to move along the $K_{ij}$
axis. Conversely, if experiments determine the actual values of
$\Delta m^2_{ij}$ and $K_{ij}$, we will gain information on both the
ratio between gauge- and gravity-mediated contributions and the
approximate horizontal symmetry that determines the flavor structure.

The information on $\Delta m^2$ probes for us the mechanism by which
supersymmetry breaking is mediated. In particular, it can determine
the mass scale of the messenger fields. It is important in this
context, however, to have complimentary information about the scale of
$m_{3/2}$ in order to test whether the $F$-term that is relevant for
gauge mediation is the same as the one that is relevant for gravity
mediation (or smaller). This information can be extracted from the
lifetime of the NLSP. (Recall that in our framework the gravitino is
the LSP.)

The information on $K^M_{ij}$, on the other hand, can give us guidance
about the way in which the SM flavor puzzle is solved. For example, if
an approximate horizontal symmetry $H$ is at work in structuring the
SM Yukawa matrices, then measurements of $K^M_{ij}$ give additional
new information about this symmetry as well as the charge assignments
$H(\Phi)$ of the various fields. In particular, while the size of the
Yukawa couplings depends on charge differences between left-handed and
right-handed field, such as $H(L_i)-H(E_j)$, those of the slepton
mass-squared matrices depend on independent combinations, such as
$H(L_i)-H(L_j)$ and $H(E_i)-H(E_j)$.  In the example models discussed
here, then, flavor information from the LHC sheds light on the same
flavor parameters that determine the SM fermion masses, yielding
additional constraints without additional degrees of freedom.  This
information is therefore likely to exclude many flavor models, and, in
some cases, would provide stringent constraints that could lead us to
compelling resolutions of both the SM and new physics flavor puzzles.

%%%%%%%%%%%%%%%%%%
\begin{acknowledgments}
YS thanks S.~Tarem and S.~Bressler for discussions.
The work of JLF, YN, and YS is supported in part by the United
States-Israel Binational Science Foundation (BSF) grant
No.~2006071.
The work of JLF is supported in part by NSF Grants Nos.~PHY--0239817
and PHY--0653656, NASA Grant No.~NNG05GG44G, and the Alfred P.~Sloan
Foundation. The research of YN is supported in part by the Israel
Science Foundation (ISF), the German-Israeli Foundation for Scientific
Research and Development (GIF), and the Minerva Foundation. The
research of YS is supported in part by the Israel Science Foundation
(ISF) under grant 1155/07.
\end{acknowledgments}

\appendix*

\section{Is Pure Alignment Viable?}

Model A in Section \ref{sec:nonmfv} employs only alignment to solve
the supersymmetric lepton flavor problem. One may ask whether a
situation where slepton masses are within reach of the LHC and there
is no degeneracy between them is worth thinking about in view of
constraints on the squark sector. In this Appendix, we would like to
make two points:
\begin{enumerate}
\item Combining the experimental constraints from $K-\bar K$ mixing
  and $D-\bar D$ mixing requires some level of degeneracy in the squark
  sector~\cite{Nir:2007xn,Ciuchini:2007cw,Nir:2007ac}.
\item Yet, it is possible to have models where the dominant
  contributions to the soft breaking terms are gravity-mediated and
  there is no degeneracy in the slepton sector.
\end{enumerate}

To make the first point, we focus our attention on the contributions
of the first two generations of squark doublets to $K-\bar K$ mixing
and to $D-\bar D$ mixing. The mass-squared matrices for these squarks
have the following form:
\begin{eqnarray}
M^2_{\tilde U_L}&=&\tilde m^2_{Q_L}+\left(\frac12 -\frac23
  \sin^2\theta_W\right)m_Z^2\cos2\beta+M_U M_U^\dagger \ ,\no\\
M^2_{\tilde D_L}&=&\tilde m^2_{Q_L}-\left(\frac12 -\frac13
  \sin^2\theta_W\right)m_Z^2\cos2\beta+M_D M_D^\dagger \ .
\end{eqnarray}
Here, $\tilde m^2_{Q_L}$ is the $2\times2$ hermitian matrix of soft
supersymmetry breaking terms. It does not break SU(2)$_L$ and
consequently it is common to $M^2_{\tilde U_L}$ and $M^2_{\tilde
D_L}$. The contribution that is proportional to $m_Z^2$ does break
SU(2)$_L$, but it respects the flavor SU(2)$_Q$ symmetry. Finally, the
contributions that are proportional to the quark mass matrices $M_U$
and $M_D$ break both SU(2)$_L$ and SU(2)$_Q$. Assuming that the mass
scale of squark doublets is in the range of 300 GeV to 1 TeV, the
contributions from $\tilde m^2_{Q_L}$ dominate over the second terms
by order $10-100$ and over the third terms by order $10^5-10^6$. This
situation leads to the following consequences:
\begin{enumerate}
\item The average squark mass $m_{\tilde q_L}$ is the same for
  left-handed up and down squarks to an accuracy that is better than
  ${\cal O}(0.1)$.
\item The mass-squared difference between the first two squark-doublet
  generations $\Delta m^2_{\tilde q_L}$ is the same in the up and down
  sectors to an accuracy that is better than ${\cal O}(10^{-5})$.
\item If the splitting between the diagonal elements of $\tilde
  m_{Q_L}$ is larger than ${\cal O}(m_c^2)$, then the mixing angles
  between the first two left-handed quark-squark generation fulfill,
  to an accuracy that is better than ${\cal O}(10^{-5})$, the
  following relation:
\begin{equation}\label{udcabibbo}
\sin\theta_{\tilde u_L}-\sin\theta_{\tilde d_L}=\sin\theta_c \ ,
\end{equation}
  where $\sin\theta_c=0.23$ is the Cabibbo angle.
\end{enumerate}

Thus, the constraints from $K-\bar K$ mixing (assuming that the
relevant phase in the supersymmetric mixing matrix is $\agt 0.1$) and
from $D-\bar D$ mixing read as follows (we take the gluino mass to be
comparable to the squark mass):
\begin{eqnarray}
\frac{1~\tev}{m_{\tilde q_L}}\frac{\Delta m^2_{\tilde
    q_L}}{m^2_{\tilde q_L}}\sin\theta_{\tilde d_L}&\leq&0.01 \ ,\no\\
\frac{1~\tev}{m_{\tilde q_L}}\frac{\Delta m^2_{\tilde
    q_L}}{m^2_{\tilde q_L}}\sin\theta_{\tilde u_L}&\leq&0.10 \ .
\end{eqnarray}
If we assume that these squarks are within the reach of the LHC,
$m_{\tilde q_L}\alt 1~\tev$, and that there is no degeneracy at all
between the first two generations of squark doublets, $\Delta
m^2_{\tilde q_L}/m^2_{\tilde q_L} \sim 1$, then the supersymmetric
flavor suppression must come entirely from alignment,
\begin{equation}\label{bouudmix}
|\sin\theta_{\tilde d_L}|\leq0.01 \ , \quad
|\sin\theta_{\tilde u_L}|\leq0.1 \ .
\end{equation}
Such a situation is, however, inconsistent with the constraint of
\eqref{udcabibbo}. We conclude that {\em if the first two generations
of squark doublets are lighter than TeV, they must be approximately
degenerate}. The minimal level of degeneracy can be derived by setting
$m_{\tilde q_L}=1~\tev$, $|\sin\theta_{\tilde d_L}|\approx0.02$, and
$|\sin\theta_{\tilde u_L}|\approx0.21$:
\begin{equation}
\label{sqdeg}
\frac{m_{\tilde q_{L2}}-m_{\tilde q_{L1}}}{m_{\tilde q_{L2}}+m_{\tilde
      q_{L1}}}\alt 0.12 \ .
\end{equation}
Hence our first point above: we know from experiment that there must
be some level of degeneracy in the squark sector.\footnote{Our
  argument holds barring accidental, fine-tuned cancelations between
  various independent supersymmetric contributions to the mixing, or
  between the supersymmetric and SM contributions.}

The second point has to do with the quantitative strength of the bound
given in \eqref{sqdeg}. Even if the squark spectrum is entirely
non-degenerate at a high scale, where supersymmetry breaking is
mediated to the supersymmetric SM, it is expected to be approximately
degenerate at low-energy. The reason is that renormalization group
evolution (RGE) introduces a universal contribution to the squark
masses-squared that is of order $7m_{\tilde g}^2$. Thus, if at the
high scale the squark masses are comparable to the gluino masses (and,
indeed, they are in the GMSB framework), then squark degeneracy at the
level of 10\% is unavoidable at low energy. Since this is the
presently required level of degeneracy, it is quite possible that
gravity-mediated contributions are comparable to (or even larger than)
the gauge-mediated ones~\cite{Nir:2002ah}. In this context, future
experimental information on the size of squark mass splittings will be
very significant, as it can distinguish between RGE and flavor-blind
mediation as the source of squark degeneracy.

The large universal contribution to squark masses is related to the
large QCD coupling. Leptons, however, have no strong interactions.
Since they experience only weak and electromagnetic interactions, the
universal RGE effects are correspondingly smaller. Indeed, the RGE of
slepton masses involves a universal contribution of order $0.3
m_{\tilde W}^2$. If at the high scale the slepton masses are
comparable to the Wino or Bino masses (and, indeed, they are in the
GMSB framework), then the absence of degeneracy at the high scale
(that is, dominance of gravity mediation) would lead to absence of
degeneracy at the low scale as well.

It is important to realize that a model-independent proof that there
must be degeneracy, similar to the one that comes from combining
\eqsref{udcabibbo}{bouudmix} in the squark sector, cannot be achieved
for the slepton sector. The reason is that, while the charged lepton
decays constrain $\sin\theta_{\tilde\ell_L}$, there is no constraint
whatsoever on $\sin\theta_{\tilde\nu_L}$. To obtain such a constraint
one need to experimentally bound processes involving external neutrino
mass eigenstates, \eg, $\nu_2\to\nu_1\gamma$, but such processes are
presently inaccessible to experiments (and, very likely, will remain
so).

The conclusion is that there cannot be a direct, model-independent
argument that suppression of supersymmetric lepton flavor violation
cannot come entirely from alignment. An indirect argument may arise in
the future, if the $D-\bar D$ mixing constraint becomes significantly
stronger, thus requiring squark degeneracy that is substantially
stronger than that given in \eqref{sqdeg}. In such a case, when it
will become unlikely that the degeneracy is a consequence of RGE only,
a flavor-blind mechanism to mediate supersymmetry breaking will be
required, suggesting that sleptons are also quasi-degenerate.

%%%%%%%%%%%%%%%%%%%%%%%%%%%%%%%%%%%%%%%%%%%%%%%%%%%%%%

%%%%%%%%%%%%%%%%%%%%%%%%%%%%%%%%%%%%%%%%%%%%%%%%%%%%%%


\begin{thebibliography}{99}
%%%%%%%%%%%%%%%%%%%%%%%%%%%%%%%%%%%%%%%%%%%%%%%%%%%%%%

%\cite{Froggatt:1978nt}
\bibitem{Froggatt:1978nt}
  C.~D.~Froggatt and H.~B.~Nielsen,
  %``Hierarchy Of Quark Masses, Cabibbo Angles And CP Violation,''
  Nucl.\ Phys.\  B {\bf 147}, 277 (1979).
  %%CITATION = NUPHA,B147,277;%%

\bibitem{Nir:2007xn}
  For a recent review, see
  Y.~Nir,
  %``Probing new physics with flavor physics (and probing flavor physics with
  %new physics),''
  arXiv:0708.1872 [hep-ph].
  %%CITATION = ARXIV:0708.1872;%%

\bibitem{Cheng:1997sq}
  H.~C.~Cheng, J.~L.~Feng and N.~Polonsky,
  %``Super-oblique corrections and non-decoupling of supersymmetry breaking,''
  Phys.\ Rev.\  D {\bf 56}, 6875 (1997)
  [arXiv:hep-ph/9706438];
  %%CITATION = PHRVA,D56,6875;%%.
%\bibitem{Cheng:1997vy}
%  H.~C.~Cheng, J.~L.~Feng and N.~Polonsky,
  %``Signatures of multi-TeV scale particles in supersymmetric theories,''
  Phys.\ Rev.\  D {\bf 57}, 152 (1998)
  [arXiv:hep-ph/9706476];
  %%CITATION = PHRVA,D57,152;%%
%\bibitem{Katz:1998br}
  E.~Katz, L.~Randall and S.~f.~Su,
  %``Supersymmetric partners of oblique corrections,''
  Nucl.\ Phys.\  B {\bf 536}, 3 (1998)
  [arXiv:hep-ph/9801416];
  %%CITATION = NUPHA,B536,3;%%
%\bibitem{Kiyoura:1998yt}
  S.~Kiyoura, M.~M.~Nojiri, D.~M.~Pierce and Y.~Yamada,
  %``Radiative corrections to a supersymmetric relation: A new approach,''
  Phys.\ Rev.\  D {\bf 58}, 075002 (1998)
  [arXiv:hep-ph/9803210];
  %%CITATION = PHRVA,D58,075002;%%
%\bibitem{Mahanta:1998hx}
  U.~Mahanta,
  %``Violation of SUSY equivalence in triple gauge boson and gaugino
  %couplings,''
  Phys.\ Rev.\  D {\bf 59}, 015017 (1999)
  [arXiv:hep-ph/9810344].
  %%CITATION = PHRVA,D59,015017;%%

\bibitem{Grossman:2007bd}
  Y.~Grossman, Y.~Nir, J.~Thaler, T.~Volansky and J.~Zupan,
  %``Probing Minimal Flavor Violation at the LHC,''
  arXiv:0706.1845 [hep-ph].
  %%CITATION = ARXIV:0706.1845;%%

\bibitem{Ciuchini:2007ha}
  M.~Ciuchini, A.~Masiero, P.~Paradisi, L.~Silvestrini, S.~K.~Vempati
and O.~Vives,
  %``Soft SUSY breaking grand unification: Leptons vs quarks on the flavor
  %playground,''
  arXiv:hep-ph/0702144.
  %%CITATION = HEP-PH/0702144;%%


\bibitem{Hayasaka:2007vc}
  K.~Hayasaka {\it et al.}  [Belle Collaboration],
  %``New search for tau --> mu gamma and tau --> e gamma decays at Belle,''
  arXiv:0705.0650 [hep-ex].
  %%CITATION = ARXIV:0705.0650;%%

\bibitem{D'Ambrosio:2002ex}
  G.~D'Ambrosio, G.~F.~Giudice, G.~Isidori and A.~Strumia,
  %``Minimal flavour violation: An effective field theory approach,''
  Nucl.\ Phys.\  B {\bf 645}, 155 (2002)
  [arXiv:hep-ph/0207036].
  %%CITATION = NUPHA,B645,155;%%

\bibitem{Cirigliano:2005ck}
  V.~Cirigliano, B.~Grinstein, G.~Isidori and M.~B.~Wise,
  %``Minimal flavor violation in the lepton sector,''
  Nucl.\ Phys.\  B {\bf 728}, 121 (2005)
  [arXiv:hep-ph/0507001].
  %%CITATION = NUPHA,B728,121;%%

\bibitem{Feng:2007sx}
  J.~L.~Feng, B.~T.~Smith and F.~Takayama,
  %``Goldilocks Supersymmetry,''
  arXiv:0709.0297 [hep-ph].
  %%CITATION = ARXIV:0709.0297;%%

\bibitem{Nir:1993mx}
  Y.~Nir and N.~Seiberg,
  %``Should squarks be degenerate?,''
  Phys.\ Lett.\  B {\bf 309}, 337 (1993)
  [arXiv:hep-ph/9304307].
  %%CITATION = PHLTA,B309,337;%%

\bibitem{Grossman:1995hk}
  Y.~Grossman and Y.~Nir,
  %``Lepton Mass Matrix Models,''
  Nucl.\ Phys.\  B {\bf 448}, 30 (1995)
  [arXiv:hep-ph/9502418].
  %%CITATION = NUPHA,B448,30;%%

\bibitem{Feng:1999wt}
  J.~L.~Feng, Y.~Nir and Y.~Shadmi,
  %``Neutrino parameters, Abelian flavor symmetries, and charged lepton flavor
  %violation,''
  Phys.\ Rev.\  D {\bf 61}, 113005 (2000)
  [arXiv:hep-ph/9911370].
  %%CITATION = PHRVA,D61,113005;%%

\bibitem{Feng:1997zr}
  J.~L.~Feng and T.~Moroi,
  %``Tevatron signatures of long-lived charged sleptons in gauge-mediated
  %supersymmetry breaking models,''
  Phys.\ Rev.\  D {\bf 58}, 035001 (1998)
  [arXiv:hep-ph/9712499].
  %%CITATION = PHRVA,D58,035001;%%

\bibitem{ArkaniHamed:1996au}
  N.~Arkani-Hamed, H.~C.~Cheng, J.~L.~Feng and L.~J.~Hall,
  %``Probing Lepton Flavor Violation at Future Colliders,''
  Phys.\ Rev.\ Lett.\  {\bf 77}, 1937 (1996)
  [arXiv:hep-ph/9603431].
  %%CITATION = PRLTA,77,1937;%%

\bibitem{ArkaniHamed:1997km}
  N.~Arkani-Hamed, J.~L.~Feng, L.~J.~Hall and H.~C.~Cheng,
  %``CP violation from slepton oscillations at the LHC and NLC,''
  Nucl.\ Phys.\  B {\bf 505}, 3 (1997)
  [arXiv:hep-ph/9704205].
  %%CITATION = NUPHA,B505,3;%%

\bibitem{Bityukov:1997ck}
  S.~I.~Bityukov and N.~V.~Krasnikov,
  %``The search for sleptons and lepton-flavor-number violation at LHC
  % (CMS),''
  Phys.\ Atom.\ Nucl.\  {\bf 62}, 1213 (1999)
  [Yad.\ Fiz.\  {\bf 62}, 1288 (1999)]
  [arXiv:hep-ph/9712358].
  %%CITATION = YAFIA,62,1288;%%

\bibitem{Agashe:1999bm}
  K.~Agashe and M.~Graesser,
  %``Signals of supersymmetric lepton flavor violation at the LHC,''
  Phys.\ Rev.\  D {\bf 61}, 075008 (2000)
  [arXiv:hep-ph/9904422].
  %%CITATION = PHRVA,D61,075008;%%

\bibitem{Hinchliffe:2000np}
  I.~Hinchliffe and F.~E.~Paige,
  %``Lepton flavor violation at the LHC,''
  Phys.\ Rev.\  D {\bf 63}, 115006 (2001)
  [arXiv:hep-ph/0010086].
  %%CITATION = PHRVA,D63,115006;%%

\bibitem{Hisano:2002iy}
  J.~Hisano, R.~Kitano and M.~M.~Nojiri,
  %``Slepton oscillation at Large Hadron Collider,''
  Phys.\ Rev.\  D {\bf 65}, 116002 (2002)
  [arXiv:hep-ph/0202129].
  %%CITATION = PHRVA,D65,116002;%%

\bibitem{Feng:2004yi}
  J.~L.~Feng and B.~T.~Smith,
  %``Slepton trapping at the Large Hadron and International Linear
  %Colliders,''
  Phys.\ Rev.\  D {\bf 71}, 015004 (2005)
  [arXiv:hep-ph/0409278].
  %%CITATION = PHRVA,D71,015004;%%

\bibitem{Hamaguchi:2004df}
  K.~Hamaguchi, Y.~Kuno, T.~Nakaya and M.~M.~Nojiri,
  %``A study of late decaying charged particles at future colliders,''
  Phys.\ Rev.\  D {\bf 70}, 115007 (2004)
  [arXiv:hep-ph/0409248].
  %%CITATION = PHRVA,D70,115007;%%

\bibitem{De Roeck:2005bw}
  A.~De Roeck, J.~R.~Ellis, F.~Gianotti, F.~Moortgat, K.~A.~Olive and
  L.~Pape,
  %``Supersymmetric benchmarks with non-universal scalar masses or
  %gravitino dark matter,''
  Eur.\ Phys.\ J.\  C {\bf 49}, 1041 (2007)
  [arXiv:hep-ph/0508198].
  %%CITATION = EPHJA,C49,1041;%%

\bibitem{Buchmuller:2004rq}
  W.~Buchmuller, K.~Hamaguchi, M.~Ratz and T.~Yanagida,
  %``Supergravity at colliders,''
  Phys.\ Lett.\  B {\bf 588}, 90 (2004)
  [arXiv:hep-ph/0402179].
  %%CITATION = PHLTA,B588,90;%%

\bibitem{Feng:2004gn}
  J.~L.~Feng, A.~Rajaraman and F.~Takayama,
  %``Probing gravitational interactions of elementary particles,''
  Int.\ J.\ Mod.\ Phys.\  D {\bf 13}, 2355 (2004)
  [arXiv:hep-th/0405248].
  %%CITATION = IMPAE,D13,2355;%%

\bibitem{Hamaguchi:2004ne}
  K.~Hamaguchi and A.~Ibarra,
  %``Probing lepton flavour violation in slepton NLSP scenarios,''
  JHEP {\bf 0502}, 028 (2005)
  [arXiv:hep-ph/0412229].
  %%CITATION = JHEPA,0502,028;%%

\bibitem{inprogress}
  J.L. Feng, S. French, C.G. Lester, Y. Nir and Y. Shadmi,
  work in progress.

\bibitem{Bartl:2005yy}
  A.~Bartl, K.~Hidaka, K.~Hohenwarter-Sodek, T.~Kernreiter,
W.~Majerotto and W.~Porod,
  %``Test of lepton flavour violation at LHC,''
  Eur.\ Phys.\ J.\  C {\bf 46}, 783 (2006)
  [arXiv:hep-ph/0510074].
  %%CITATION = EPHJA,C46,783;%%

\bibitem{Deppisch:2007rm}
  F.~Deppisch,
  %``Lepton Flavor Violation at the LHC,''
  arXiv:0710.2525 [hep-ph].
  %%CITATION = ARXIV:0710.2525;%%

\bibitem{Assamagan:2002kf}
  K.~A.~Assamagan, A.~Deandrea and P.~A.~Delsart,
  %``Search for the lepton flavor violating decay A0/H0 --> tau+- mu-+ at
  %hadron colliders,''
  Phys.\ Rev.\  D {\bf 67}, 035001 (2003)
  [arXiv:hep-ph/0207302].
  %%CITATION = PHRVA,D67,035001;%%

\bibitem{Frixione:2002ik}
  S.~Frixione and B.~R.~Webber,
  %``Matching NLO QCD computations and parton shower simulations,''
  JHEP {\bf 0206}, 029 (2002)
  [arXiv:hep-ph/0204244].
  %%CITATION = JHEPA,0206,029;%%

\bibitem{Frixione:2003ei}
  S.~Frixione, P.~Nason and B.~R.~Webber,
  %``Matching NLO QCD and parton showers in heavy flavour production,''
  JHEP {\bf 0308}, 007 (2003)
  [arXiv:hep-ph/0305252].
  %%CITATION = JHEPA,0308,007;%%

\bibitem{Mangano:2002ea}
  M.~L.~Mangano, M.~Moretti, F.~Piccinini, R.~Pittau and A.~D.~Polosa,
  %``ALPGEN, a generator for hard multiparton processes in hadronic
  %collisions,''
  JHEP {\bf 0307}, 001 (2003)
  [arXiv:hep-ph/0206293].
  %%CITATION = JHEPA,0307,001;%%

%\bibitem{atlasphysicstdr1}
%  The ATLAS Collaboration,
%  ``ATLAS Detector and Physics Performance Technical Design Report 1'',
%  CERN-LHCC-99-014, ATLAS-TDR-14 (1999).

%\bibitem{atlasphysicstdr2}
%  The ATLAS Collaboration,
%  ``ATLAS Detector and Physics Performance Technical Design Report 2'',
%  CERN-LHCC-99-015 ATLAS-TDR-15 (1999).

\bibitem{Ellis:2006vu}
  J.~R.~Ellis, A.~R.~Raklev and O.~K.~Oye,
  %``Gravitino dark matter scenarios with massive metastable charged
  %sparticles at the LHC,''
  JHEP {\bf 0610}, 061 (2006)
  [arXiv:hep-ph/0607261].
  %%CITATION = JHEPA,0610,061;%%

\bibitem{Polesello:683824}
G.~Polesello and A.~Rimoldi,
``Reconstruction of quasi-stable charged sleptons in the
ATLAS Muon Spectrometer'',
CERN,
ATL-MUON-99-006,
(1999).

\bibitem{Ambrosanio:2000ik}
  S.~Ambrosanio, B.~Mele, S.~Petrarca, G.~Polesello and A.~Rimoldi,
  %``Measuring the SUSY breaking scale at the LHC in the slepton NLSP scenario
  %of GMSB models,''
  JHEP {\bf 0101}, 014 (2001)
  [arXiv:hep-ph/0010081].
  %%CITATION = JHEPA,0101,014;%%

\bibitem{Zalewski:2007up}
  P.~Zalewski,
  %``Search for GMSB NLSPs at LHC,''
  arXiv:0710.2647 [hep-ph].
  %%CITATION = ARXIV:0710.2647;%%

\bibitem{Bressler:2007gk}
  S.~Bressler  [ATLAS Collaboration],
  %``R-Hadron and long lived particle searches at the LHC,''
  arXiv:0710.2111 [hep-ex];
  %%CITATION = ARXIV:0710.2111;%%
S.~Tarem, S.~Bressler, H.~Nomoto, A.~Dimattia [ATLAS Collaboration],
in preparation.

\bibitem{Paige:1996nx}
  F.~E.~Paige,
  %``Determining SUSY particle masses at LHC,''
{\it In the Proceedings of 1996 DPF / DPB Summer Study on
New Directions for High-Energy Physics (Snowmass 96), Snowmass,
Colorado, 25 Jun - 12 Jul 1996, pp SUP114}
  [arXiv:hep-ph/9609373].
  %%CITATION = ECONF,C960625,SUP114;%%

\bibitem{Hinchliffe:1998ys}
  I.~Hinchliffe and F.~E.~Paige,
  %``Measurements in gauge mediated SUSY breaking models at LHC,''
  Phys.\ Rev.\  D {\bf 60}, 095002 (1999)
  [arXiv:hep-ph/9812233].
  %%CITATION = PHRVA,D60,095002;%%

\bibitem{Bachacou:1999zb}
  H.~Bachacou, I.~Hinchliffe and F.~E.~Paige,
  %``Measurements of masses in SUGRA models at LHC,''
  Phys.\ Rev.\  D {\bf 62}, 015009 (2000)
  [arXiv:hep-ph/9907518].
  %%CITATION = PHRVA,D62,015009;%%

\bibitem{Allanach:2000kt}
  B.~C.~Allanach, C.~G.~Lester, M.~A.~Parker and B.~R.~Webber,
  %``Measuring sparticle masses in non-universal string inspired models at the
  %LHC,''
  JHEP {\bf 0009}, 004 (2000)
  [arXiv:hep-ph/0007009].
  %%CITATION = JHEPA,0009,004;%%

\bibitem{Allanach:2004ub}
  B.~C.~Allanach {\it et al.}  [Beyond the Standard Model Working Group],
  %``Les Houches 'Physics at TeV Colliders 2003' Beyond the Standard Model
  %Working Group: Summary report,''
  arXiv:hep-ph/0402295.
  %%CITATION = HEP-PH/0402295;%%

\bibitem{Nojiri:2003tu}
  M.~M.~Nojiri, G.~Polesello and D.~R.~Tovey,
  %``Proposal for a new reconstruction technique for SUSY processes at the
  %LHC,''
  arXiv:hep-ph/0312317.
  %%CITATION = HEP-PH/0312317;%%

\bibitem{Gjelsten:2005aw}
  B.~K.~Gjelsten, D.~J.~Miller and P.~Osland,
  %``Measurement of the gluino mass via cascade decays for SPS 1a,''
  JHEP {\bf 0506}, 015 (2005)
  [arXiv:hep-ph/0501033].
  %%CITATION = JHEPA,0506,015;%%

\bibitem{Miller:2005zp}
  D.~J.~Miller, P.~Osland and A.~R.~Raklev,
  %``Invariant mass distributions in cascade decays,''
  JHEP {\bf 0603}, 034 (2006)
  [arXiv:hep-ph/0510356].
  %%CITATION = JHEPA,0603,034;%%

\bibitem{Lester:2006cf}
  C.~G.~Lester, M.~A.~Parker and M.~J.~White,
  %``Three body kinematic endpoints in SUSY models with non-universal Higgs
  %masses,''
  JHEP {\bf 0710}, 051 (2007)
  [arXiv:hep-ph/0609298].
  %%CITATION = JHEPA,0710,051;%%



%\cite{Ciuchini:2007cw}
\bibitem{Ciuchini:2007cw}
  M.~Ciuchini, E.~Franco, D.~Guadagnoli, V.~Lubicz, M.~Pierini,
V.~Porretti and L.~Silvestrini,
  %``D-Dbar mixing and new physics: general considerations and constraints
%on the MSSM,''
  Phys.\ Lett.\  B {\bf 655}, 162 (2007)
  [arXiv:hep-ph/0703204].
  %%CITATION = PHLTA,B655,162;%%

%\cite{Nir:2007ac}
\bibitem{Nir:2007ac}
  Y.~Nir,
  %``Lessons from BaBar and Belle measurements of D-Dbar mixing
%parameters,''
  JHEP {\bf 0705}, 102 (2007)
  [arXiv:hep-ph/0703235].

\bibitem{Nir:2002ah}
  Y.~Nir and G.~Raz,
  %``Quark squark alignment revisited,''
  Phys.\ Rev.\  D {\bf 66}, 035007 (2002)
  [arXiv:hep-ph/0206064].
  %%CITATION = PHRVA,D66,035007;%%

%%%%%%%%%%%%%%%%%%%%%%%%%%%%%%%%%%%%%%%%%%%%%%%%%%%%%%
\end{thebibliography}
\end{document}